\documentstyle[twocolumn,prl,aps,epsf]{revtex}
\begin{document}
\draft
%\preprint{}

\title{NaV$_2$O$_5$ as a quarter-filled ladder compound} 

\author{Holger Smolinski, Claudius Gros, Werner Weber} 
\address{   Institut f\"ur
 Physik, Universit\"at Dortmund, 44221 Dortmund, Germany.}
\author{Ulrich Peuchert, Georg Roth} 
\address{Institut f\"ur Kristallographie, RWTH Aachen, 52056 
         Aachen, Germany}
%  FAX:  (0241) 8888 184
\author{Michael Weiden und Christoph Geibel}
\address{MPI f\"ur Chemische Physik fester Stoffe, 
        TH-Darmstadt, 64289 Darmstadt, Germany}

\date{\today}

\maketitle

\begin{abstract}
A new X-ray diffraction study of the one-dimensional
spin-Peierls compound $\alpha^\prime$-NaV$_2$O$_5$
reveals a centrosymmetric (Pmmn) crystal structure 
with one type of V site, contrary to the
previously postulated non-centrosymmetric P2$_1$mn
structure with two types of V sites (V$^{+4}$ and V$^{+5}$).
Density functional calculations indicate that
NaV$_2$O$_5$ is a quarter-filled ladder compound with 
the spins carried by V-O-V molecular orbitals 
on the rungs of the ladder.
Estimates of the charge-transfer gap and the exchange coupling 
agree well with experiment and
explain the insulating behavior of NaV$_2$O$_5$ and
its magnetic properties.
\end{abstract}


%%%%%%%%%%%%%%%%%%%%%%%%%%%%%%%%%%%%%%%%%%%%%%%%%%%%%%%%%%%%%%%%%%%%%%%%
%%%%%%%%%%%%%%%%%%%%%%%%%%%%%%%%%%%%%%%%%%%%%%%%%%%%%%%%%%%%%%%%%%%%%%%%
\centerline{\hfill}

\paragraph*{Introduction}
Two classes of quasi one-dimensional
compounds with spin-gaps have been investigated 
intensively in the last few years, 
ladder systems like SrCu$_2$O$_3$ and
Sr$_{14}$Cu$_{24}$O$_{41}$ \cite{Azuma,Uehara}
and spin-Peierls compounds like
CuGeO$_3$ and $\alpha^\prime$-NaV$_2$O$_5$ \cite{Hase,Isobe}. 
For even-leg ladder systems the spin-gap is
structurally induced and present at all
temperatures (for an overview see \cite{Dagotto}).
The possibility of superconductivity in doped
ladder systems has been discussed \cite{Dagotto}
and found in Sr$_2$Ca$_{12}$Cu$_{24}$O$_{41}$ under
pressure \cite{Uehara}. A spin-Peierls system
undergoes a lattice instability at $T_{SP}$ and
for $T<T_{SP}$ the system dimerizes and a spin-gap
opens. Here we propose that the spin-Peierls
compound NaV$_2$O$_5$ is at the same time a quarter-filled
ladder system, in contrast to the previous notion
which assumed NaV$_2$O$_5$ to be made up of weakly
coupled pairs of V$^{+4}$ and V$^{+5}$ chains \cite{Isobe,Galy_Carpy}.
Our proposition is based on a re-determination of the
crystal structure of NaV$_2$O$_5$ by X-ray diffraction
and on density-functional calculations. 
Mapping of the density functional results on Hubbard and
Heisenberg models yields values for the model parameters
which explain readily the insulating
behavior of NaV$_2$O$_5$ and its the magnetic properties.
Our results also show that NaV$_2$O$_5$ and
CaV$_2$O$_5$ are isostructural and consequently
establish CaV$_2$O$_5$ as a half-filled ladder system.

\begin{figure}[bth]
   \epsfxsize=0.48\textwidth
   \centerline{\epsffile{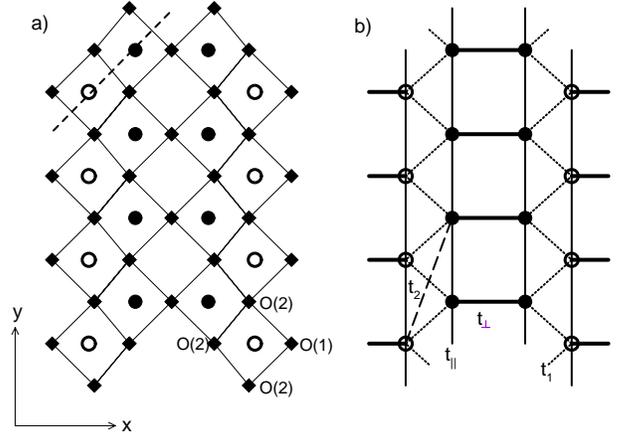}}
   \centerline{\parbox{0.48\textwidth}{\caption{\label{Fig1}
   a) Crystal structure of NaV$_2$O$_5$. The diamonds denote
   the position of the oxygen O(1)- and O(2)-ions which form
   the basal quadrangles of the VO$_5$ pyramids oriented
   in $\pm z$ directions. Also shown  are the positions of the
   equivalent V1/V2-atoms, located above/below the basal plan
   (empty/filled circles). Not indicated are the apex O(3)
   and the Na positions. 
   The dashed line indicates the intersection of the
   plane for the charge density plot of Fig.~\protect\ref{Fig3}.
   b) Hopping matrix elements of our effective V(d$_{xy}$)
   Hamiltonian.
   }}}
   \end{figure}

\paragraph*{Crystal structure}
The crystal structure of $\alpha^\prime$-NaV$_2$O$_5$ 
consists of double chains of
edge-sharing distorted tetragonal VO$_5$-pyramids
running along the orthorhombic b-axis,
which are linked together via common corners 
of the pyramids to form sheets. These in turn
are stacked upon each other along c with no direct 
V-O-V links.
(see Fig.~\ref{Fig1} and \cite{Isobe,Galy_Carpy}).
The Na atoms are located between these sheets.
In their original papers 
Galy {\it et al.} and Carpy {\it et al.} 
\cite{Galy_Carpy} proposed the 
non-centrosymmetric space group C$^7_{2v}$-P2$_1$mn for this 
mixed valence compound (on average V$^{+4.5}$).
After the discovery of a spin-Peierls 
transition in $\alpha^\prime$-NaV$_2$O$_5$ at $T_{SP}=34~{\rm K}$
\cite{Isobe}, it has been argued 
that the charge ordering would lead to a magnetic 
decoupling of adjacent double chains \cite{Isobe,Fujii}
and would be responsible for the one-dimensional character 
of this compound observed in magnetic susceptibility
measurements \cite{Isobe,Mila}.
%%%%%%%%%%%%%%%%%%%%%%%%%%%%%%%%%%%%%%%%%%%%%%%%%%%%%%%%%%%%%%%%%
%%%%%%%%%%%%%%%%%%%%%%%%%%%%%%%%%%%%%%%%%%%%%%%%%%%%%%%%%%%%%%
%
%
%%%%%%%%%%%%%%%%%%%%%%%%%%%%%%%%%%%%%%%%%%%%%%%%%%%%%%%%%%%%%%%%%
%%%%%%%%%%%%%%%%%%%%%%%%%%%%%%%%%%%%%%%%%%%%%%%%%%%%%%%%%%%%%%

Our re-determination of the structure by single crystal X-ray 
diffraction at room temperature, however, shows very
clearly that the structure of NaV$_2$O$_5$ 
is in fact centrosymmetric (D$_{2h}^{13}$-Pmmn)
with only one distinct V-position and 3 
instead of 5 inequivalent oxygen atoms. 
NaV$_2$O$_5$ is consequently isostructural to
CaV$_2$O$_5$ \cite{Bouloux,Onoda}.
The relevant experimental details and results are summarized in 
Table \ref{Table1}.
The topology of the structure remains essentially unchanged with
respect to previous results \cite{Galy_Carpy}. 
The possibility for long range charge ordering, however, 
is lost in the higher symmetry group.

%%%%%%%%%%%%%%%%%%%%%%%%%%%%%%%%%%%%%%%%%%%%%%%%%%%%%%%%%%%%%%%%%
%
   \begin{figure}[tbh]
   \epsfxsize=0.48\textwidth
   \centerline{\epsffile{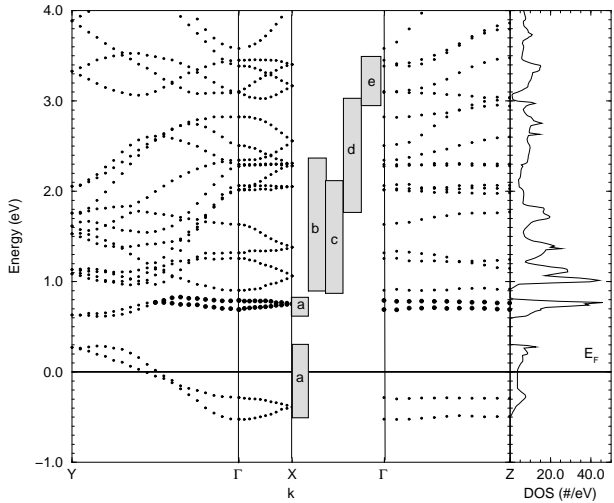}}
   \medskip
   \centerline{\parbox{0.48\textwidth}{\caption{\label{Fig2}
   V-d bands and density of states (DOS) as obtained from DFT.
   Also shown is a term scheme of the V-d orbitals, derived from
   an analysis of partial density of states.
   Here a,b,c,d, e denote respectively the
   d$_{xy}$, d$_{yz}$, d$_{zx}$, d$_{x^2-y^2}$
   and d$_{3z^2-r^2}$ orbitals. The filled dots in
   the second-lowest subband pair denote
   the range in k-space of O(1)-p$_y$ admixture
   in the anti-bonding d$_{xy}$ band-pair.
   }}}
   \end{figure}
%
%%%%%%%%%%%%%%%%%%%%%%%%%%%%%%%%%%%%%%%%%%%%%%%%%%%%%%%%%%%%%%%%%

Attempts to refine the structure in the 
acentric space group P2$_1$mn resulted in unphysical 
anisotropic displacement parameters and strong correlations 
between positional as well as displacement parameters of atoms 
which are equivalent in the centrosymmetric space group. 
Despite the larger number of variables in the acentric space group, 
the reliability values did not improve at all. 
Attempts to refine under the assumption of merohedral twinning 
(incoherent superposition of "inversion-twins") did not succeed either.
Finally, a thorough inspection of a large number of centrosymmetric 
pairs of reflections (so-called Friedel-equivalents) did 
not reveal any significant differences which could be indicative 
of a missing center of symmetry. 
These results are based on a number of measurements on various 
flux-grown crystals \cite{Weiden} from different batches, at two 
temperatures (293~K, 173~K), and with different wavelengths 
(MoK$\alpha$, AgK$\alpha$). Also, it was verified by 
Raman and susceptibility measurements \cite{Weiden}
that the crystals used for the diffraction experiments indeed
exhibit a sharp spin-Peierls transition at 34~K. The
data presented in Table \ref{Table1} are
for the homogeneous phase ($T>T_{SP}$), the crystal
structure for $T<T_{SP}$ has yet to be determined in
detail \cite{Fujii}.

\paragraph*{Band structure}
Based on the new crystal structure (Table \ref{Table1})
we have calculated the energy bands of NaV$_2$O$_5$
within density functional theory (DFT)\cite{relaxed}.
Thereby we have employed the 
full-potential linearized augmented plane
wave code WIEN97 \cite{WIEN}. We have
treated the exchange-correlation part by using
the generalized gradient approximation \cite{Perdew}. 
Also, local orbitals have been included for a better
description of the semicore states
(of Na-2s, Na-2p, V-3s, V-3p and O-1s).

The V-3d energy bands span a width of $\approx5~{\rm eV }$
(see Fig.~\ref{Fig2}). The bottom of the 3d bands
is separated by $\approx3~{\rm eV}$ from the top of the
O-2p band manifold. The sequence of 3d subband-splittings
is in accordance with estimates by ligand field theory
from recent $^{51}$V NMR data \cite{Ohama}.

\begin{figure}[htb]
    \epsfxsize=0.48\textwidth
    \centerline{\epsffile{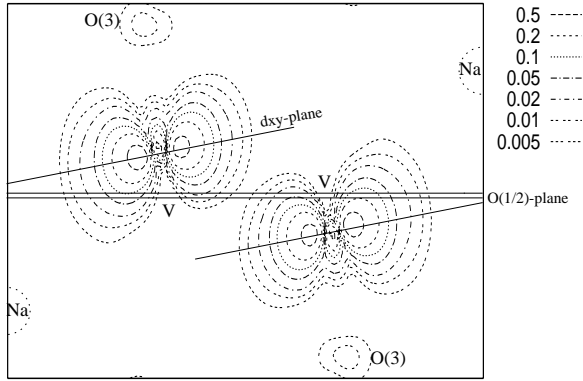}}
    \medskip
    \centerline{\parbox{0.48\textwidth}{\caption{\label{Fig3}
    Partial charge density (e/\AA$^3$) of the V-d bands in the
    plane indicated Fig.~\protect\ref{Fig1}. This plane
    contains the Na sites and approaches closely the
    V1, V2 and apex O(3) sites.
    The planes of the tilted d$_{xy}$-orbital 
    point $\pi$-like towards the O(1). 
    }}}
    \end{figure}

The four lowest-lying d bands predominantly exhibit
V-d$_{xy}$ character (see Fig.~\ref{Fig2}). Actually, the 
d$_{xy}$ orbital planes are somewhat tilted around the
b-axis towards the respective O(1) positions above or below 
the centers of the V-O-V rungs (see Fig.~\ref{Fig3}).
The four d$_{xy}$ bands are split into two pairs of subbands,
separated by $\approx 0.5-1{\rm eV}$. All bands exhibit significant
dispersion along $\Gamma$-Y, to a lesser extend along $\Gamma$-X,
but hardly any dispersion along $\Gamma$-Z 
(see Fig.~\ref{Fig2}). 

Analysis \cite{analysis} of the DFT band states and mapping of the bands
on those of tight-binding models yield the following
result (model I): 
Bonding-type molecular orbital states, made up
by the d$_{xy}$-orbitals of a V-O-V rung, coupled via the
hopping term $t_\perp\approx-0.38~{\rm eV}$, form the lower
pairs of subbands. Their dispersion along $\Gamma$-Y is
produced by $t_\parallel\approx-0.17~{\rm eV}$, their
splitting at $\Gamma$, the small dispersion along
$\Gamma$-X and the band-crossing along $\Gamma$-Y result
from the small inter-ladder hoppings 
$t_1\approx0.012~{\rm eV}$ and $t_2\approx0.03~{\rm eV}$
(terms see Fig.~\ref{Fig1} b)).
The upper pair of subbands consists of the corresponding
anti-bonding molecular orbitals.

In order to understand the
microscopic origin of the effectiv hopping parameter
$t_\perp$, $t_\parallel$, $t_1$ and  $t_2$,
we have extended the tight-binding model
by including the p$_x$ and p$_y$ orbitals of the basal plane
O(1) and O(2) atoms (model II). We find strong contributions
to $t_\perp$ and $t_\parallel$ (of model I)
by the indirect coupling of the V-d$_{xy}$ orbitals via the
O-p, predominantly of pd$\pi$-type. Further contributions
to $t_\perp$ and $t_\parallel$ arise from residual three-center 
d-d terms involving the anionic oxygen potentials.
The ratio $t_\perp/t_\parallel\approx3.1$ is related to the
tilt of the V-d$_{xy}$ orbitals, which reduces the 
V-d$_{xy}$ --- O(2)-p$_x$ coupling along b and 
increases the V-d$_{xy}$ --- O(1)-p$_y$ coupling.
The residual direct d-d couplings are affected
correspondingly by the tilt.

Further, the indirect contribution to $t_1$ via the
V1-O(2)-V2 path is small due to a Goodenough-Kanamori-Anderson 
type interference effect in the
(almost square-like) quadrangels of V1-O(2)-V2-O(2), which is
similar to that discussed in CuGeO$_3$ \cite{Mattheiss}.
The residual direct contributions to $t_1$ are affected
by a compensation effect between two- and
three-center contributions of opposite signs.

The sign of $t_\parallel$ is derived from a symmetry analysis
of the DFT eigenfunctions at the $\Gamma$-point.
Admixture of O(1)-p$_y$ orbitals occurs only in the
antibonding states of the V-O-V rung
(upper subband pair). The p$_y$-admixture
is strongest near $\Gamma$, but disappears completely near Y
(see Fig.~\ref{Fig2}). The latter feature points to a
strong O(1)-O(1) pp$\sigma$ hopping ($\approx1$ eV) along b.

%%%%%%%%%%%%%%%%%%%%%%%%%%%%%%%%%%%%%%%%%%%%%%%%%%%%%%%%%%%%%%%%%
% 
%
%%%%%%%%%%%%%%%%%%%%%%%%%%%%%%%%%%%%%%%%%%%%%%%%%%%%%%%%%%%%%%%%%

The model resulting from our tight-binding analysis
is that of a system of weakly
coupled V-ladders with 0.5 electrons per site (quarter filled).
The band states are superpositions of d$_{xy}$ - d$_{xy}$
molecular orbitals of bonding and antibonding type.
The pair of antibonding bands is empty and the the pair
of bonding bands is half-filled. If we ignore the
small inter-ladder couplings $t_1$ and $t_2$, the d$_{xy}$
bond orbitals, which show up as the lowest subband-pair in
Fig.~\ref{Fig2}, form a pair of half-filled one-dimensional
bands. Let us note, that this picture is robust and does not
depend on the details of the tight-binding analysis. Inclusion
of other intra-ladder hopping parameters and O(1)-O(2) hopping
matrix elements leads only to renormalized tight-binding
parameters, but not to a fundamental modification of
the model discussed above. Details will be discussed elsewhere.

Correlation effects will lead to a charge gap and to the
experimentally observed insulating behavior.
Only one gapless spin-excitation branch remains \cite{Balents}.
These statements hold both in the cases of weak and
of strong electron-electron interaction $U$.
We have estimated the value of U by a DFT-calculation
where we have doped a fractional number of
extra electrons. Though the doping is compensated on the
Na sites, all additional electrons enter the V-d$_{xy}$ bands. The
extra charge density causes a shift of all V-d bands with respect
to the O-p bands. In Hartree-Fock approximation for our model I
with on-site interaction U,the shift 
$\Delta E$ of the one-particle energies
caused by the extra charge density $\Delta n$ is given by
\cite{Vielsack}
\[
\Delta E = U \Delta n/ 2.
\]
For $\Delta n=0.1$ we find a energy shift 
$\Delta E = 0.14$ eV. We estimate 
$U\approx 2.8$ eV and a strong-coupling picture
is therefore appropriate for NaV$_2$O$_5$. 
We may thus regard NaV$_2$O$_5$
as being built up from
weakly coupled spin-1/2 chains with each
spin located in a bonding V-O-V molecular 
wavefunction. Below $T_{SP}$ these chains
of V-O-V spins dimerize and a spin-gap
opens.
%%%%%%%%%%%%%%%%%%%%%%%%%%%%%%%%%%%%%%%%%%%%%%%%%%%%%%%%%%%%%%%%%
%
   \begin{figure}[htb]
   \epsfxsize=0.48\textwidth
   \centerline{\setlength{\unitlength}{0.012500in}%
\begingroup\makeatletter\ifx\SetFigFont\undefined
% extract first six characters in \fmtname
\def\x#1#2#3#4#5#6#7\relax{\def\x{#1#2#3#4#5#6}}%
\expandafter\x\fmtname xxxxxx\relax \def\y{splain}%
\ifx\x\y   % LaTeX or SliTeX?
\gdef\SetFigFont#1#2#3{%
  \ifnum #1<17\tiny\else \ifnum #1<20\small\else
  \ifnum #1<24\normalsize\else \ifnum #1<29\large\else
  \ifnum #1<34\Large\else \ifnum #1<41\LARGE\else
     \huge\fi\fi\fi\fi\fi\fi
  \csname #3\endcsname}%
\else
\gdef\SetFigFont#1#2#3{\begingroup
  \count@#1\relax \ifnum 25<\count@\count@25\fi
  \def\x{\endgroup\@setsize\SetFigFont{#2pt}}%
  \expandafter\x
    \csname \romannumeral\the\count@ pt\expandafter\endcsname
    \csname @\romannumeral\the\count@ pt\endcsname
  \csname #3\endcsname}%
\fi
\fi\endgroup
%\begin{picture}(225,145)(5,690)
\begin{picture}(225,115)(5,710)
\thicklines
%%%%%%%%%%%%%%%%%%%%%%%%%%%%%%%%%%%%%%%%%
\put( 30,710){\line( 1, 0){ 40}}
\put( 43,707){\makebox(0,0)[lb]{\smash{\SetFigFont{12}{14.4}{rm}
$\uparrow$}}}
\put(  5,708){\makebox(0,0)[lb]{\smash{\SetFigFont{12}{14.4}{rm}
-t$_\perp$}}}
%%%%%%%%%%%%%%%%%%%%%%%%%%%%%%%%%%%%%%%%%
\put( 30,760){\line( 1, 0){ 40}}
\put( 15,758){\makebox(0,0)[lb]{\smash{\SetFigFont{12}{14.4}{rm}0}}}
%%%%%%%%%%%%%%%%%%%%%%%%%%%%%%%%%%%%%%%%%
\put( 30,810){\line( 1, 0){ 40}}
\put( 43,807){\makebox(0,0)[lb]{\smash{\SetFigFont{12}{14.4}{rm}
$\downarrow$}}}
\put(  5,808){\makebox(0,0)[lb]{\smash{\SetFigFont{12}{14.4}{rm}
t$_\perp$}}}
%%%%%%%%%%%%%%%%%%%%%%%%%%%%%%%%%%%%%%%%%
\put( 80,760){\line( 1, 0){ 40}}
\put( 90,757){\makebox(0,0)[lb]{\smash{\SetFigFont{12}{14.4}{rm}
$\uparrow\downarrow$}}}
%%%%%%%%%%%%%%%%%%%%%%%%%%%%%%%%%%%%%%%%%
\put(145,757){\line( 1, 0){ 40}}
\put(145,760){\line( 1, 0){ 40}}
\put(145,763){\line( 1, 0){ 40}}
\put(145,730){\line( 1, 0){ 40}}
\put(190,757){\makebox(0,0)[lb]{\smash{\SetFigFont{12}{14.4}{rm}
0}}}
\put(190,727){\makebox(0,0)[lb]{\smash{\SetFigFont{12}{14.4}{rm}
-J$_\perp$}}}
%%%%%%%%%%%%%%%%%%%%%%%%%%%%%%%%%%%%%%%%%
\multiput(123,759)(3.75000,-7.00000){5}{
\makebox(0.1111,0.7778){\SetFigFont{5}{6}{rm}\large.}}
\multiput(123,759)(7.50000,0.00000){3}{
\makebox(0.1111,0.7778) {\SetFigFont{5}{6}{rm}\large.}}
\end{picture}}
   \medskip\medskip\medskip
   \centerline{\parbox{0.48\textwidth}{\caption{\label{Fig4}
   Term scheme of an empty, single and doubly-occupied
   V-O-V orbital representing an isolated rung of the ladder 
   in the  $U\gg t_\perp$ limit.
   Given are the respective energies, $J_\perp\approx4\,t_\perp^2/U$. 
   }}}
   \end{figure}
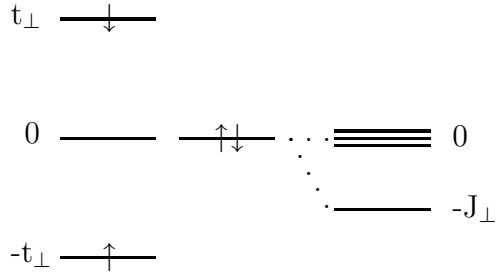
%
%%%%%%%%%%%%%%%%%%%%%%%%%%%%%%%%%%%%%%%%%%%%%%%%%%%%%%%%%%%%%%%%%

We now proceed to estimate the charge-transfer gap
$E_c$ and the antiferromagnetic coupling
$J_\parallel$. We neglect the inter-ladder
couplings $t_1$ and $t_2$. 
Since $t_\perp\sim 3t_\parallel$ we 
may use perturbation-theory in $t_\parallel/t_\perp$.
The one- and two-particle states of an isolated
V-O-V rung are depicted in Fig.~\ref{Fig4}. The bonding
and antibonding one-particle states have the
energies $-t_\perp$ and $+t_\perp$ respectively.
The exchange integral along a rung is
$J_\perp=4\,t_\perp^2/U\approx0.41~{\rm eV}$.
In the ground state of the
ladder all bonding states are filled and the first
excited charge-transfer state is given by one empty
and one doubly occupied rung with energy
\[
E_c =2t_\perp - J_\perp\approx 0.71 {\rm eV}.
\]
This value for the charge-transfer agrees well with
the $0.6-0.7$ eV observed in optical absorption
spectra \cite{Golubchik}. The exchange coupling
$J_\parallel$
between adjacent V-O-V molecular spins can be estimated
by standard perturbation theory in $t_\parallel/t_\perp$
and is given by
\[
J_\parallel ={2t_\parallel^2\over E_c} \approx 80~{\rm meV},
\]
which corresponds to 930~K. The exchange 
integral for NaV$_2$O$_5$ has been estimated
to be 560-700~K \cite{Isobe,note_Horsch}. 
DFT therefore overestimates $J_\parallel$ 
somewhat. 

\paragraph*{Outlook}
The above results indicate that $\alpha^\prime$-NaV$_2$O$_5$ 
may be the first known {\it quarter-filled} ladder compound
\cite{Dagotto}. It is possible to
dope charge carriers into the insulating quarter-filled
state. One way
is to introduce Na defects, Na$_x$V$_2$O$_5$, 
as the $\alpha^\prime$-phase is stable
for $0.7<x<1.0$ \cite{Hardy}. Alternatively one may
consider the Ca substitution, Na$_{1-y}$Ca$_y$V$_2$O$_5$,
since our results establish NaV$_2$O$_5$
and CaV$_2$O$_5$ to be isostructural \cite{Bouloux,Onoda}.
Varying $y\in[0,1]$ would then allow to increase
the carrier concentration continuously until a
(highly anisotropic) half-filled ladder compound 
is obtained for $y=1$.
Note, that a spin-gap has been measured for
CaV$_2$O$_5$ \cite{Onoda,Iwase}.

\paragraph*{Conclusions}
So far $\alpha^\prime$-NaV$_2$O$_5$ has been considered 
as an inorganic spin-Peierls compound
\cite{Isobe,Fujii}, where
V$^{+4}$ and V$^{+5}$ ions are ordered in
parallel chains\cite{Galy_Carpy}.
In this letter we present evidence that
NaV$_2$O$_5$ is in fact a quarter-filled ladder
system, consisting of equivalent V atoms, 
and that NaV$_2$O$_5$ is 
isostructural to CaV$_2$O$_5$. Our results
establish CaV$_2$O$_5$ to be a half-filled
ladder compound and thus explain the
observed spin-gap in CaV$_2$O$_5$.
Our arguments are
based on the crystallographic re-examination
of NaV$_2$O$_5$ by X-ray diffraction and energy-band
calculations. We find the crystal structure to
be centrosymmetric with only one equivalent V ion. 
This result does not
allow for spontaneous charge disproportionation
V$^{+4}$ - V$^{+5}$. A tight-binding analysis
of the band structure leads to a one-dimensional Heisenberg
model in the low-energy sector with a spin of 1/2
per rung of the ladder. These spins are not
attached to a single V ion, but to a V-O-V
molecular orbital. Our estimates for the
charge-transfer gap and the exchange coupling
agree with experiment.

We acknowledge useful discussions with
P. Lemmens, R. Noack, A. Kluemper and M. Braden.

Note added in proof: A recent V-NMR study
also finds only one equivalent V-site in
NaV$_2$O$_5$ above $T_{SP}$ \cite{NMR}, as
does a crystal structure redetermination \cite{Schnering}.
The possibility of a molecular spin-state has also been
discussed recently \cite{Horsch98}.

%%%%%%%%%%%%%%%%%%%%%%%%%%%
%%%%%%%%%%%%%%%%%%%%%%%%%%%

%\kern-10pt

%%%%%%%%%%%%%%%%%%%%%%%%%%%%%%%%%%%%%%%%%%%%%%%%%%%%%%%%%%%%%%%%%
%%%%%%%%%%%%%%%%%%%%%%%%%%%%%%%%%%%%%%%%%%%%%%%%%%%%%%%%%%%%%%%%%
%
\begin{table}[htb]
  \begin{tabular}{ccccc}
    & $x$     &  $y$  & $z$     & u(eq) \\
V   & 4021(1) &  2500 & 3920(1) &  7(1) \\
Na  & 2500    & -2500 & 8593(1) & 17(1) \\
O(1)& 2500    &  2500 & 5193(2) &  9(1) \\
O(2)& 5731(1) &  2500 & 4877(1) &  9(1) \\
O(3)& 3854(1) &  2500 &  578(1) & 15(1)\\
  \end{tabular}
   \medskip
  \caption{\label{Table1} 
\underline{Data collection parameters}:
NaV$_2$O$_5$, Pmmn, compare Fig.~\protect\ref{Fig1}: 
  a=11.316(4)\AA, b=3.611(1)\AA, c=4.797(2)\AA, 
z=2, T=293(2)K, $\lambda$=0.561\AA, 
$\sin\theta/\lambda|_{\rm max.}=1.365\AA^{-1}$, 
full sphere, numerical absorption correction, 
R$_{\rm int}$= 0.052, 2279 indep.\ data, 27 parameters, 
R$_{\rm F}$=0.0238, wR$_{\rm F}^2$=0.0483 for 
${\rm I}>2\sigma_{\rm I}$.\\
\underline{Atomic coordinates} ($\times10^4$) and equivalent 
isotropic displacement parameters ($\AA^2\times10^3$), see
Table above. The two additional O(2) depicted in
Fig.~\protect\ref{Fig1} at
(0.4269,0.75,0.5123) and
(0.4269,-0.25,0.5123) are generated by symmetry.  \\
\underline{Bond lengths (\AA)}: 
V-O(1): 1.8263(6), 
V-O(2): 1.9161(5) (2x), 
V-O(2): 1.9887(8) (1x), 
V-O(3): 1.6144(9), 
Na-O: 2.5353(9) (avg.), 
shortest V-V: 3.0400(7) , shortest V-Na: 3.3537(8).
          }
\end{table}

%%%%%%%%%%%%%%%%%%%%%%%%%%%%%%%%%%%%%%%%%%%%%%%%%%%%%%%%%%%%%%%%%
%%%%%%%%%%%%%%%%%%%%%%%%%%%%%%%%%%%%%%%%%%%%%%%%%%%%%%%%%%%%%%%%%
%%%%%%%%%%%%%%%%%%%%%%%%%%%%%%%%%%%%%%%%%%%%%%%%%%%%%%%%%%%%%%%%%


\begin{thebibliography}{99}
\bibitem{relaxed} We have found, that, within DFT, the P2$_1$mn
                  structure relaxes to the Pmmn structure.
\bibitem{analysis} For the analysis we have used the dispersion of the
                   bonding states and the center of gravity of the
                   antibonding states.
\end{thebibliography}
\end{document}